\documentclass{PoS}

\title{Conserved charge fluctuations with HISQ fermions }

\ShortTitle{Conserved charge fluctuations with HISQ fermions }

\author{\speaker{Christian Schmidt} (for the BNL-Bielefeld Collaboration\thanks{Collaboration members are: A. Bazavov, H.-T. Ding, P. Hegde, O. Kaczmarek, F. Karsch, E. Laermann, 
S. Mukherjee, P. Petreczky, C. Schmidt, D. Smith, W. S\"oldner,  M. Wagner.}$\;\;$)\\
        Universit\"at Bielefeld, Fakult\"at f\"ur Physik, Postfach 100131, D-33501 Bielefeld, Germany\\
        E-mail: \email{schmidt@physik.uni-bielefeld.de}}

\abstract{We calculate cumulants of fluctuations of net-baryon number, net-electric charge and net-strangeness, in the framework of lattice regularized QCD. We use a highly improved staggered quark (HISQ) action on lattices with temporal extent of $N_\tau=6,8$ and 12 and almost physical quark masses.  By means of a Taylor expansion in various chemical potentials and under demanding both strangeness neutrality as well as the correct isospin asymmetry, we evaluate these fluctuations at conditions met in heavy ion collisions. Cumulants of net-electric charge fluctuations can, in principle, also be measured in heavy ion experiments. We therefore propose a method to extract freeze-out parameters, such as the freeze-out temperature and baryon chemical potential, based on a comparison of lattice results and experimental measurements of two different ratios of net-electric charge cumulants. As this method involves only (lattice) QCD results and experimental measurements it is model independent and, for the purpose of extracting freeze-out parameters, does not require any input from hadron resonance gas model calculations. }

\FullConference{Xth Quark Confinement and the Hadron Spectrum,\\
		October 8-12, 2012\\
		TUM Campus Garching, Munich, Germany}

\begin{document}

\section{Introduction}
\begin{figure}[htbp]
\begin{center}
\begin{minipage}{0.5\textwidth}
\begin{center}
\includegraphics[width=0.99\textwidth]{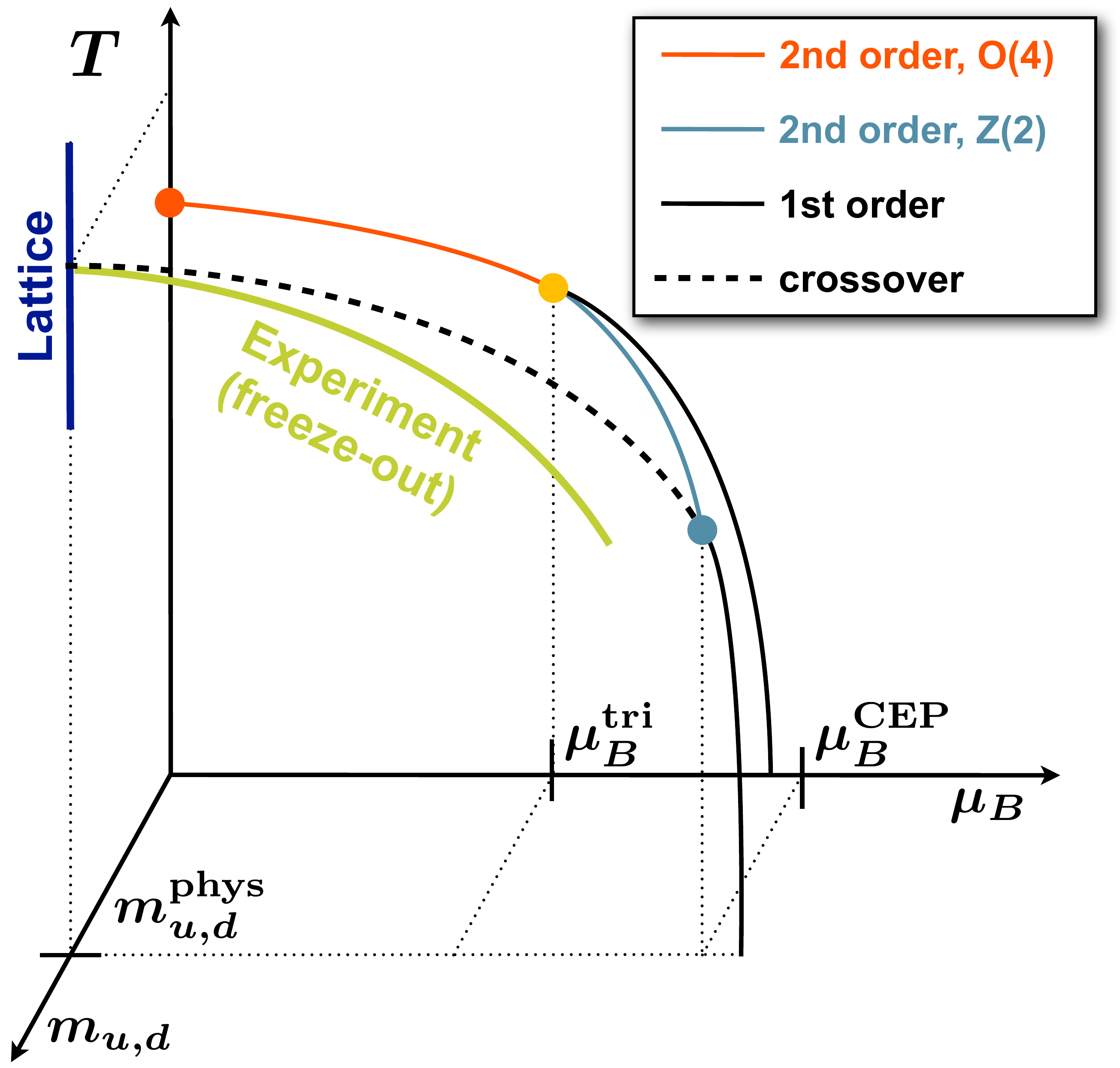}
\end{center}
\end{minipage}
\end{center}
\caption{Generic phase diagram of QCD, based on model calculations and model independent symmetry arguments.  Also indicated are the regions in the phase diagram where we are able to obtain results on fluctuation observables from lattice QCD and experiments, respectively. \label{fig:pdiag}}
\end{figure}

To obtain a detailed and quantitative understanding of the QCD phase diagram is one of the most important and outstanding problems in high energy physics. The analysis of the phase diagram is naturally connected with an investigation of the properties of strongly interacting matter at high temperature and densities and the nature of the QCD phase transition. One expects that the nature of the transition depends on the quark masses and chemical potentials. A generic phase diagram based on model calculations and model independet arguments is shown in Fig.~\ref{fig:pdiag} \cite{pdiag}.  One of the most interesting question is to clarify whether there exists a critical point in the phase diagram at physical quark masses or not.

Recently  large efforts are made in probing the QCD phase diagram with heavy ion collision. The RHIC beam energy scan program (BES) aims at a systematic scan of the QCD phase diagram. Through a variation of the center of mass energy of the two colliding ions, the fireball made of quark gluon plasma (QGP) is generated under different initial conditions, as {\it e.g.}, different initial temperatures and net-baryon densities. After its formation, it evolves on isentropic (hydrodynamic) trajectories in the phase diagram, {\it i.e.} it expands and cools. It eventually goes through the QCD transition and hadronizes. The point from whereon the abundance of the hadronic particle species do no longer change is called chemical freeze-out. The particle yields realized here are then finally measured by various detectors, only modified by particle decays. One expects that under the variation of the collision energy ($\sqrt{s_{AA}}$), the freeze-out happens along a one-dimensional curve in the phase diagram that can be parametrized by $\sqrt{s_{AA}}$ \cite{fc}. The freeze-out curve is also shown in Fig.~\ref{fig:pdiag}. One way to calculate the freeze-out parameters such as temperature, chemical potentials and volume is by performing a least-square-fit of the Hadron Resonance Gas (HRG) model to the measured particle yields. This procedure has been very successful in the past \cite{HRG}.    

Theoretically best understood is the QCD phase diagram at vanishing net-baryon density, or equivalently zero baryon chemical potential. Here lattice regularized calculations of QCD are feasible and have reached a precision level that enables controlled continuum extrapolated results with physical quark masses. Also this region is indicated in Fig.~\ref{fig:pdiag}. Unfortunately, lattice QCD calculations are not possible at non-vanishing net-baryon densities by means of standard Monte Carlo methods, due to the notorious sign problem. Nevertheless, there are possibilities to extract lattice QCD results at small but nonzero baryon chemical potentials. A straightforward approach is, {\it e.g.}, given by expanding the observables in a Taylor series around zero baryon chemical potential ($\mu_B\equiv 0$) \cite{taylor}. A method to calculate the freeze-out curve by a direct comparison of lattice results with experimental measurements has been recently proposed by us \cite{letter,qm1,qm2,xqcd}. It is based on a Taylor expansion of ratios of cumulants of net-charge fluctuations. Since it involves only first principle lattice QCD calculations it is (HRG) model independent.

\section{Cumulants of conserved charges\label{sec:fluct}}
We start with considering an expansion of the QCD partition functions -- or rather the pressure in dimensionless units -- in terms of the net-baryon, net-electric charge and net-strangeness chemical potentials, also in dimensionless units ($\hat \mu_B\equiv\mu_B/T,\hat \mu_Q\equiv \mu_Q/T,\hat \mu_S\equiv \mu_S/T$) which is given by 
\begin{equation}
\frac{p}{T^4}=\frac{1}{VT^3}\ln Z=\sum_{i,j,k}\frac{1}{i!\; j!\; k!\;}\chi_{ijk,0}^{BQS}\;{\hat \mu_B}^i \; {\hat \mu_Q}^j \; {\hat \mu_S}^k\;.
\end{equation}
We calculate the temperature dependent coefficients $\chi_{ijk,0}^{BQS}$ on the lattice at vanishing chemical potentials. On the lattice we obtain them as derivatives of the partition function (generalized susceptibilities) exactly as given by the definition of the Taylor series
\begin{equation}
\left(VT^3\right)\cdot\chi^{BQS}_{ijk,0}(T)= \left(\partial^{i+j+k}\ln Z(T,\mu_B,\mu_Q,\mu_S)\right) \left/ \left(\partial \hat\mu_B^i \partial \hat\mu_Q^j \partial \hat\mu_S^k \right)\right|_{\vec\mu=0}\;.
\label{eq:fluct1}
\end{equation} 
Interestingly, these coefficients also define cumulants of the corresponding net-charge fluctuations ($N_B, N_Q, N_S$), for the diagonal coefficients one finds 
\begin{eqnarray}
\left(VT^3\right)\cdot\chi^X_2 &=& \left<\left(\delta N_X\right)^2\right>,\\
\left(VT^3\right)\cdot\chi^X_4 &=& \left<\left(\delta N_X\right)^4\right>-3\left<\left(\delta N_X\right)^2\right>^2,\\
\left(VT^3\right)\cdot\chi^X_6 &=& \left<\left(\delta N_X\right)^6\right>-15\left<\left(\delta N_X\right)^4\right>\left<\left(\delta N_X\right)^2\right>+30\left<\left(\delta N_X\right)^2\right>^3,
\label{eq:fluct2}
\end{eqnarray}
with $X=B,Q,S$ and $\delta N_X=N_X-\left<N_X\right>$. Similar relations can also be found for the off-diagonal coefficients. The cumulants of charge fluctuations can in principal be measured in heavy ion experiments. However, as experiments have difficulties to trigger on neutrons and various strange particles, baryon number and strangeness fluctuations seem to be unfeasible. On the other hand, experiments can measure proton number fluctuations \cite{Luo}, which can not be calculated in QCD since the proton number is not a conserved quantity. Unless one finds a way to relate proton number to baryon number fluctuations \cite{Kitazawa:2012at,Bzdak:2012an} the only remaining set of cumulants that can be obtained in both experiments and lattice QCD calculations are the net-electric charge fluctuations. Nevertheless, we have to keep in mind, that the measurement of the fluctuations of a conserved charge is a difficult experimental task, as it involves a delicate tuning of the rapidity window that is accepted in the detectors \cite{Koch}. Choosing the acceptance range too large, will result in a suppressed signal as the global charge conservations forces the fluctuations to vanish in the limit of a total ($4\pi$) acceptance, on the other hand, an acceptance that is too small will loose the sensitivity to the relevant physics and/or not survive hadronization and the hadronic phase.

In lattice QCD calculations electric charge fluctuations have systematic errors that need to be controlled. Although statistical errors are much smaller than in the baryon number fluctuations, one has to work on very fine lattices in order to control the continuum limit systematically. This is due to the fact that large contributions to the electric charge fluctuations stem from the light pion sector. The pion spectrum is distorted on the lattice, when one uses the popular  staggered fermion formulation of the QCD action. For reasons of the numerical costs, staggered fermions are the most common type of fermions for thermodynamic calculations on the lattice. In the following we will use the highly improved staggered fermion (HISQ) action \cite{HISQ}, with reduces the deformation of the pion sector as much as possible. 

\begin{figure}[htbp]
\begin{center}
\begin{minipage}{0.325\textwidth}
\begin{center}
\includegraphics[width=1.0\textwidth]{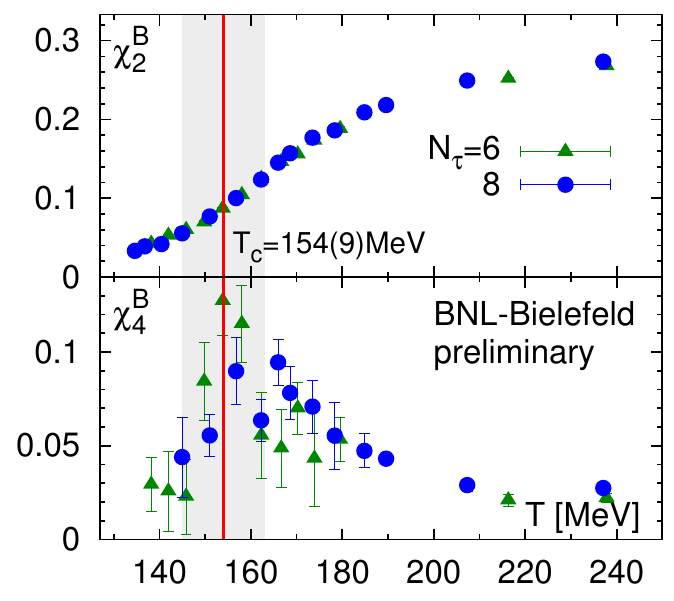}
\end{center}
\end{minipage}
\begin{minipage}{0.325\textwidth}
\begin{center}
\includegraphics[width=1.0\textwidth]{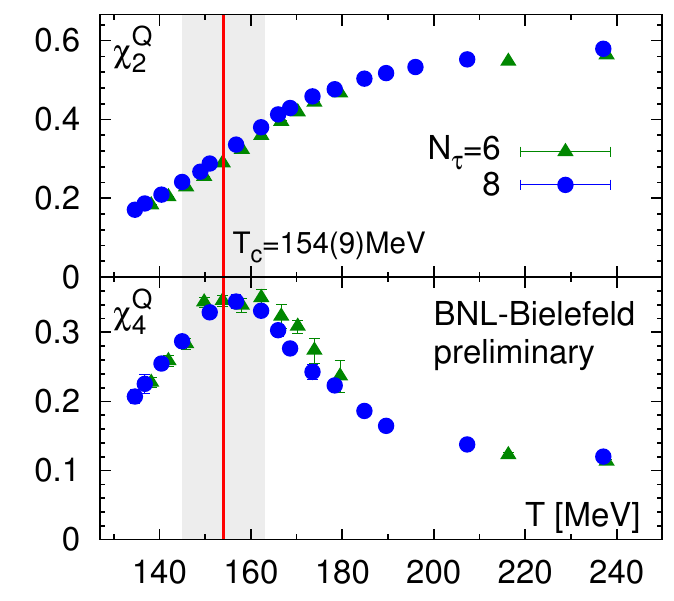}
\end{center}
\end{minipage}
\begin{minipage}{0.325\textwidth}
\begin{center}
\includegraphics[width=1.0\textwidth]{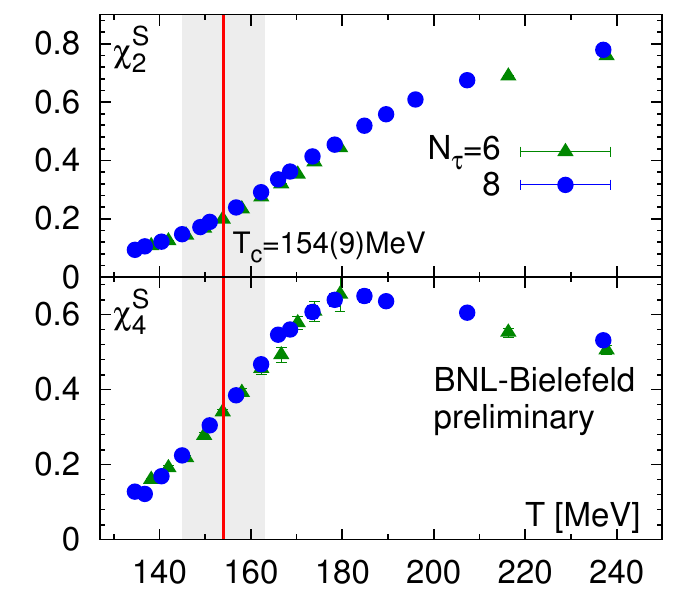}
\end{center}
\end{minipage}
\end{center}
\caption{Second and fourth order cumulants of net-baryon number fluctuations (left column), net-electric charge fluctuations (middle column) and net-strangeness fluctuations (right column). Different symbols denote different lattice spacings: $N_\tau=6$ (triangles) and $8$ (circles). The vertical bar indicates the QCD transition temperature as obtained in \cite{Tc}. \label{fig:chi}}
\label{fig:multi_X}
\end{figure}

In Fig.~\ref{fig:multi_X} we show our preliminary results of the cumulants (expansion coefficients), obtained with the HISQ action. The calculations have been done with 2 light and one heavier quark flavor and almost physical quark masses, {\it i.e.} the strange quark mass has been fixed to its physical value, whereas the light quark mass was taken to be $1/20$ of the strange quark mass.
In general we find little dependence on the lattice cutoff, which is here controlled by the number of lattice points in (Euclidean) temporal direction ($N_\tau$). In case of the $2^{\rm nd}$ order fluctuations the HotQCD Collaboration has performed a continuum extrapolation based on $N_\tau=6,8$ and $12$ lattices \cite{HotQCD}. Good agreement with HRG model results has been found for temperatures below $T\lesssim 150$ MeV.  

The structure of the cumulants is found to be consistent with the expected QCD critical behavior connected to the $O(4)$ critical point in the massless limit of the two light quark flavors (see Fig.~\ref{fig:pdiag}). From an analysis of the free energy in terms of universal scaling fields one would expect the singular part of the cumulants to behave as $\chi^X_{2n}\sim |t|^{2-n-\alpha}$ \cite{Ejiri:2005wq}, at least for $X=B$ and $Q$, where $t$ is the reduced temperature and $\alpha$ the critical exponent of the specific heat.  As $\alpha$ is small and negative ($\alpha\approx-0.15$) the $4^{\rm th}$ order fluctuations should develop a cusp in the chiral limit whereas the $6^{\rm th}$ order fluctuations are the first to diverge.

\section{Electric charge and strangeness chemical potentials\label{sec:muQS}}  
\begin{figure}[htbp]
\begin{center}
\includegraphics[width=0.32\textwidth]{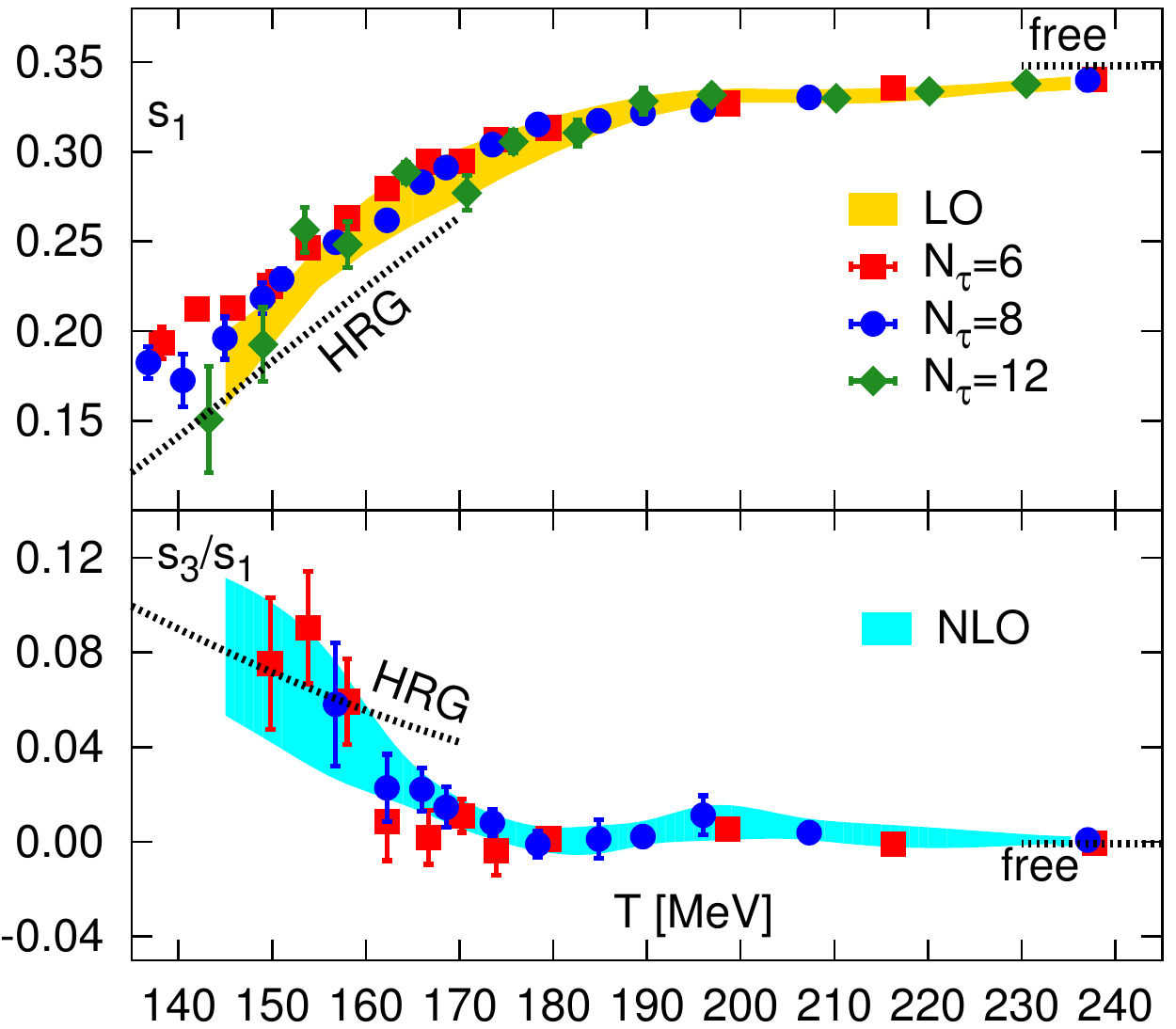} 
\includegraphics[width=0.32\textwidth]{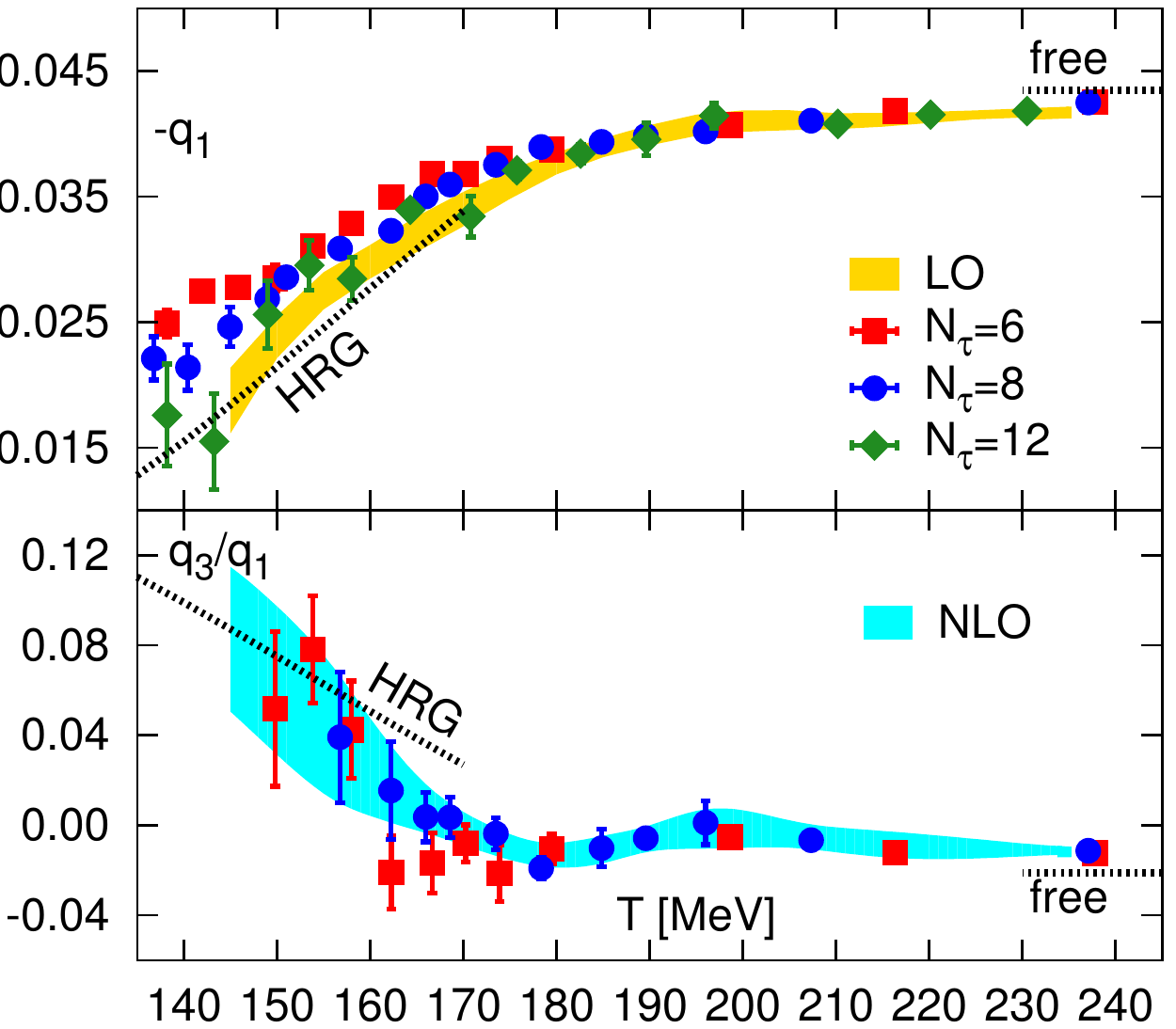} 
\includegraphics[width=0.32\textwidth]{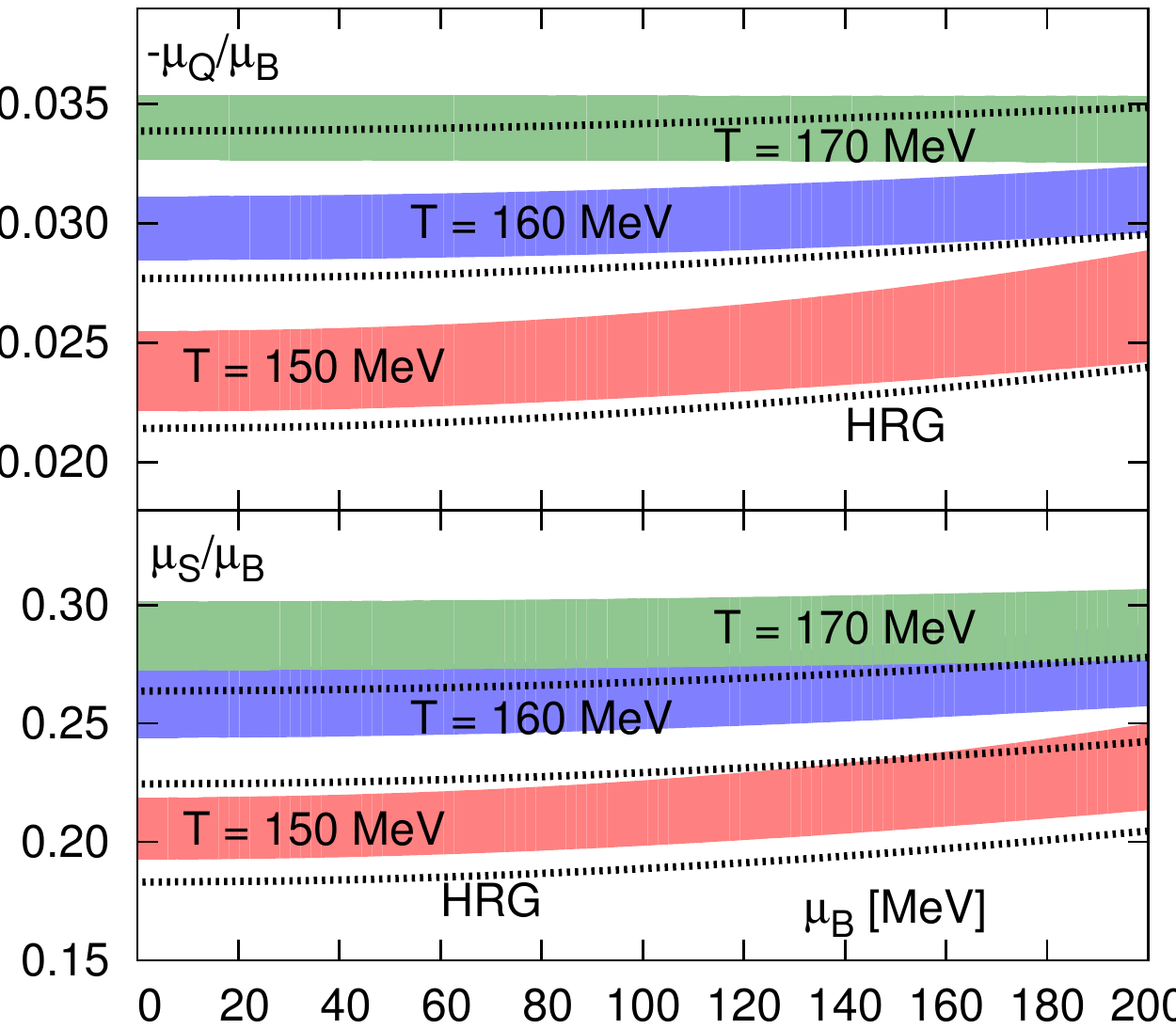} 
\caption{The leading and next-to-leading order expansion coefficients 
of the strangeness (left) and the negative of the electric charge chemical 
potentials (middle) versus temperature for $r=0.4$.
For $s_1$ and $q_1$ the LO-bands show results for the continuum 
extrapolation. For $s_3$ and $q_3$ we give an estimate for continuum 
results (NLO bands) based on spline interpolations of the $N_\tau=8$ data.
Dashed lines at low temperature are from the HRG model
and at high temperature from a massless, 3-flavor quark gas. 
The right hand panel shows NLO results for $\mu_S/\mu_B$ and
$\mu_Q/\mu_B$ as function of $\mu_B$ for three values of the 
temperature.}
\label{fig:chempot}
\end{center}
\end{figure}

In order to resemble the conditions met in heavy ion collisions as closely as possible, we demand strangeness neutrality $(\left< N_S\right>=0)$ and the correct isospin asymmetry ($r=\left< N_Q\right>/\left< N_B\right>$). These two conditions can be realized by choosing the free parameter $\hat\mu_Q$ and $\hat\mu_S$ accordingly. 
By expanding both the conditions as well as the chemical potentials in terms of $\hat \mu_B$,
\begin{equation}
\hat\mu_S=s_1(T)\hat\mu_B+s_3(T)\hat\mu_B^3+\mathcal{O}(\hat\mu_B^5),\qquad
\hat\mu_Q=q_1(T)\hat\mu_B+q_3(T)\hat\mu_B^3+\mathcal{O}(\hat\mu_B^5),
\label{eq:chemex}
\end{equation}
we can solve for the coefficients in Eq.~\ref{eq:chemex} order by order and thus enforce the conditions up to arbitrary order in $\hat\mu_B$. Or results for the leading order (LO) and next to leading order (NLO) coefficients of the $\hat\mu_Q$ and $\hat\mu_S$ series are shown in Fig.~\ref{fig:chempot} (left and middle). 
The upper panel shows the leading order (LO), whereas the lower panels show the ratio of NLO to LO coefficients. The band in the upper panels indicates the continuum extrapolation based on the $N_\tau=6,8$ and 12 data, the band in the lower panels is a continuum estimate based on spline interpolations of the $N_\tau=8$ data. We find that the NLO contributions are negligible in the high temperature region and below 10\% in the temperature interval relevant for the analysis of freeze-out conditions, {\it i.e.}, $T\approx (160 \pm 10)$ MeV. In fact, in this temperature range the leading order lattice QCD results deviate from HRG model calculations expanded to the same order by less than 15\%. Note that one can also investigate the convergence properties of the HRG model itself.  In the HRG model the NLO expansion reproduces the full HRG result for $\hat\mu_Q$ and $\hat\mu_S$ to better than 1.0\% for all values of $\hat\mu_B\lesssim 1.3$. Altogether, we thus expect that the NLO truncated QCD expansion is a good approximation to the complete QCD results for $\hat\mu_Q$ and $\hat\mu_S$ for $\mu_B \lesssim 200$ MeV.

Our results for the strangeness and electric charge chemical potentials at NLO as function of  $\mu_B$ and $T$ are shown in Fig.~\ref{fig:chempot} (right). While $\mu_S/\mu_B$ varies between 0.2 and 0.3 in the interval 150 MeV $\lesssim T \lesssim 170$ MeV, the absolute value of $\mu_Q/\mu_B$ is an order of magnitude smaller. Both ratios are almost constant for $\mu_B \lesssim 200$ MeV, which is consistent with HRG model calculations.

\section{Comparison with the experiment}
We will now construct the observables that we want to compare with the experiment in order to determine the remaining freeze-out parameters, which are the freeze-out temperature ($T^f$), the freeze-out baryon chemical potential ($\mu_B^f$) and the freeze-out volume ($V^f$). The latter one can be easily eliminated by considering ratios of cumulants as should be apparent from Eqs.~\ref{eq:fluct1} -\ref{eq:fluct2}. We are thus left with two freeze-out parameters ($T^f,\mu_B^f$), for which we need two independent observables to match with the experiment. As already discussed in Sec.~\ref{sec:fluct}, from the set of cumulants here considered, only the net-electric charge fluctuations can be determined on the lattice as well as measured in experiments. We therefore propose the following two ratios of net-electric charge fluctuations for the comparison with the experiment.
\begin{eqnarray}
R_{12}^Q &\equiv& 
\frac{M_Q}{\sigma_Q^2} =
\frac{\chi_1^Q}{\chi_2^Q}= 
\hat\mu_B \left(
R_{12}^{Q,1} + R_{12}^{Q,3}\ \hat\mu_B^2 + {\cal O}(\hat\mu_B^4)
\right)\; ,
\label{R12} \\
R_{31}^Q &\equiv& 
\frac{S_Q \sigma_Q^3}{M_Q} = 
\frac{\chi_3^Q}{\chi_1^Q}=
 R_{31}^{Q,0} + R_{31}^{Q,2}\ \hat\mu_B^2 + {\cal O}(\hat\mu_B^4)\; .
\label{R31}
\end{eqnarray}
Here we expressed the cumulant ratios also in terms of the mean value ($M$), the variance ($\sigma$) and the skewness ($S$), which characterize the shape of the net-electric charge distribution. We have again expanded these two quantities in terms of $\hat \mu_B$ and determined the LO and NLO contributions in that series (the remaining chemical potentials $\mu_Q$ and $\mu_S$ have been fixed as described in Sec.~\ref{sec:muQS}). The two ratios defined in Eq.~\ref{R12} and \ref{R31} represent the most simple choice as they only involve the evaluation of cumulants up to the $2^{\rm nd}$ and $4^{\rm th}$ order, respectively, at LO and up to the $4^{\rm th}$ and $6^{\rm th}$ order, respectively, at NLO. Morover, they are complementary in the sense that $R_{12}^{Q}$ (Eq.~\ref{R12}) is an odd function of $\hat\mu_B$, which leads to a distinct sensitivity with respect to the freeze-out chemical potential ($\mu_B^f$), whereas $R_{31}$ (Eq.~\ref{R31}) is an even function of $\hat\mu_B$, starting with a constant, which results in a much more pronounced sensitivity to the freeze-out temperature ($T^f$).

\begin{figure}[htbp]
\begin{center}
\includegraphics[width=0.32\textwidth]{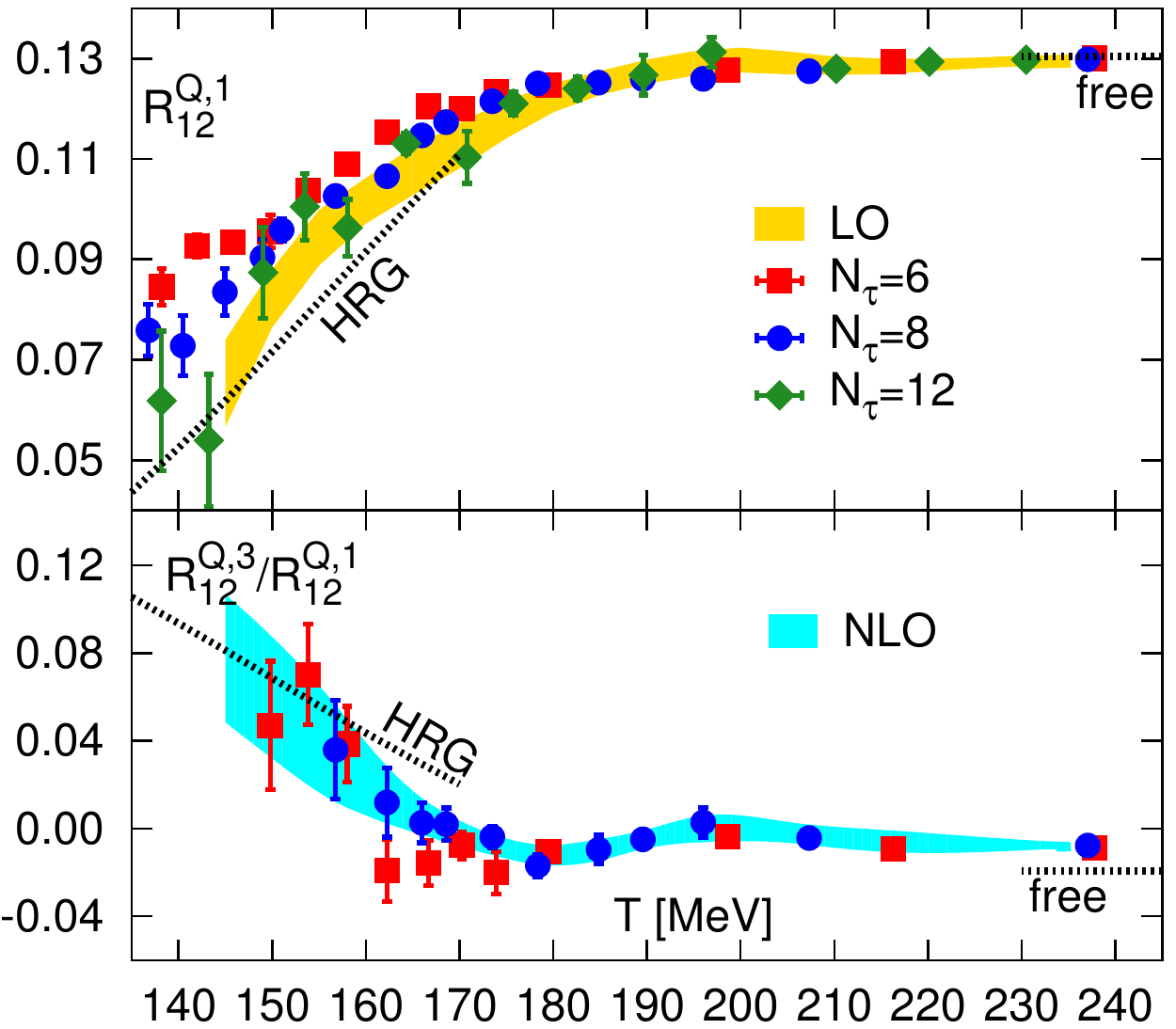}
\includegraphics[width=0.32\textwidth]{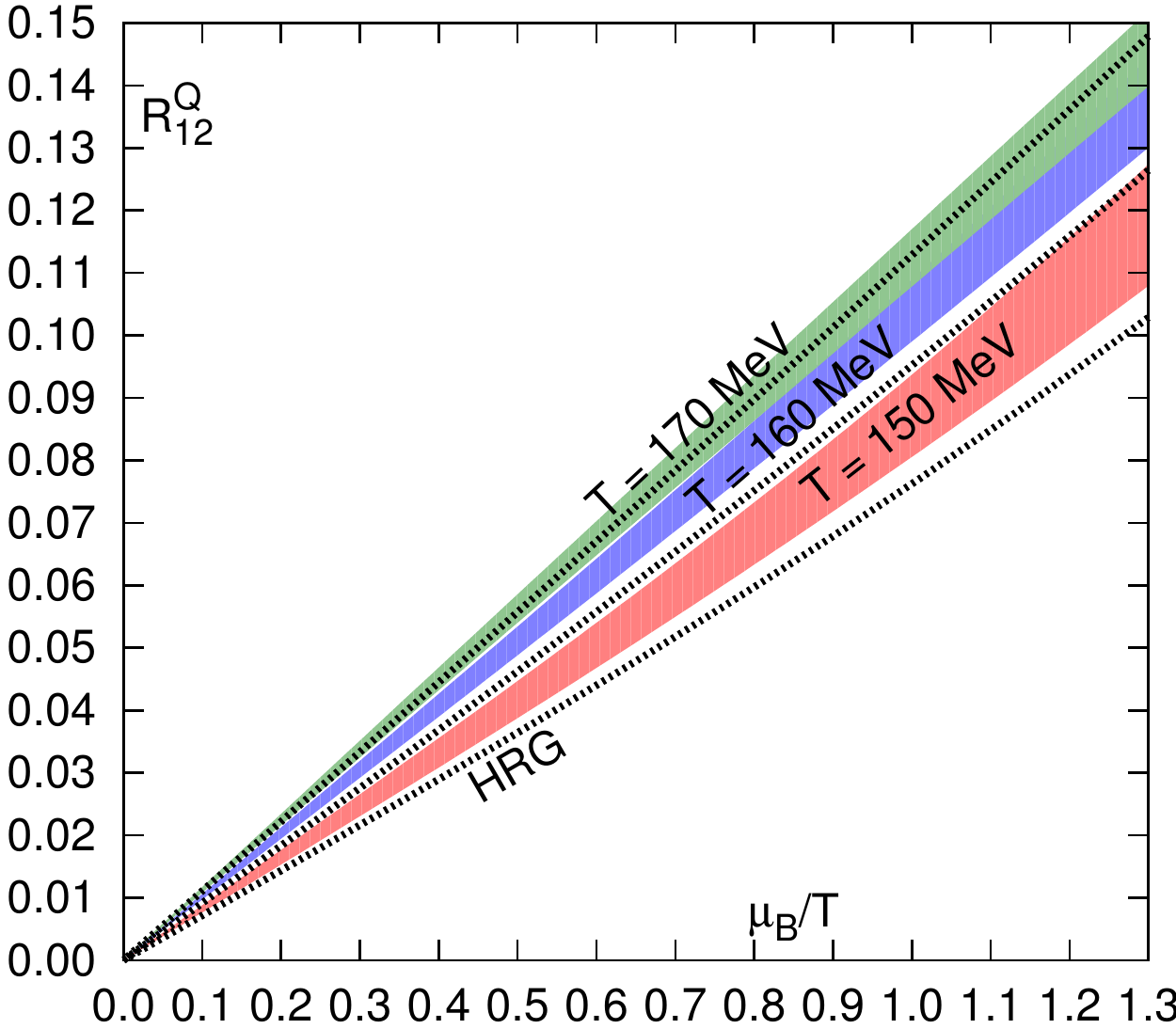}
\includegraphics[width=0.32\textwidth]{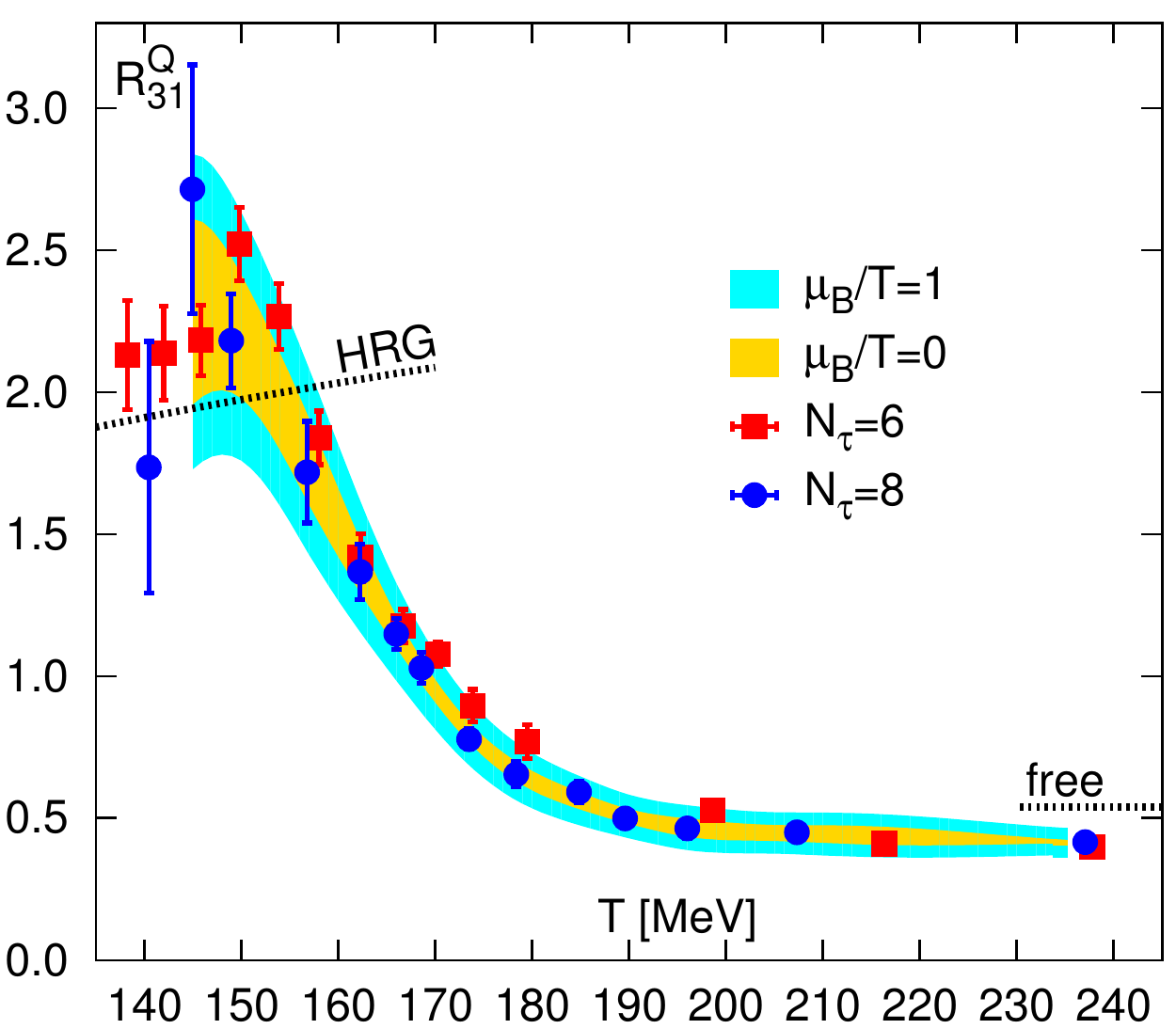}
\caption{The left panel shows LO (top) and NLO (bottom) expansion coefficients of $R_{12}^Q$ for $r=0.4$. The bands and lines are as in 
Fig.~\protect\ref{fig:chempot}(left).On the middle panel we plot $R_{12}^Q$ versus $\mu_B/T$, including the NLO contribution, for three
values of the temperature. $R_{31}^Q$ versus temperature is shown on the right panel for $\mu_B=0$. The wider band on the data set for 
$N_\tau=8$ shows an estimate of the magnitude of NLO corrections. }
\label{fig:R31}
\end{center}
\end{figure}

In Fig.~\ref{fig:R31} (left) we show our results on the LO and NLO expansion coefficients of $R_{12}^Q$. The bands on the lower and upper panel have the same meaning as in Fig.~\ref{fig:chempot} (left). 
We find that the NLO corrections to $R_{12}^Q$ are below 10\%, which makes the LO result a good approximation for a large range of $\hat\mu_B$. Systematic errors arising from the truncation of the Taylor series for $R_{12}^Q$ at NLO may again be estimated by comparing the full result in the HRG model calculation with the corresponding truncated results. Here we find for $T = (160 \pm 10)$ MeV and $\hat\mu_B\lesssim 1.3$ that the difference is less than 1.0\%. Moreover, we estimated that taste violation effects in the NLO calculation lead to systematic errors that are at most 5\% and thus will be negligible in $R_{12}^Q$ . Taylor series truncated at NLO are thus expected to give a good approximation to the full result for a wide range of baryon chemical potentials.

In Fig.~\ref{fig:R31} (middle) we show the full $\mu_B$ and $T$ dependence of $R_{12}^Q$, including the NLO contribution. Obviously the ratio $R_{12}^Q$ shows a strong sensitivity on $\mu_B$ but varies little with $T$ in the temperature range $T=(160\pm 10)$ MeV. For the determination of $(T^f,\mu_B^f)$ a second, complimentary information is needed. To this end we use the ratio $R_{31}^Q$, which is strongly dependent on $T$ but receives corrections only at $\mathcal{O}(\hat\mu_B^2)$. The leading order result for this ratio is shown in Fig.~\ref{fig:R31} (right).  Apparently this ratio shows a characteristic temperature dependence for $T\gtrsim 155$ MeV that is quite different from that of HRG model calculations. The NLO correction to this ratio vanishes in the high temperature limit and at low $T$ the HRG model also suggests small corrections. In fact, in the HRG model the LO contributions to $R_{31}^Q $ differ by less than 2\% from the exact results on the freeze-out curve for $\mu_B \lesssim 200$ MeV. The broader band in Fig.~\ref{fig:R31} (right) indicates an estimate of the NLO contribution at $\hat\mu_B=1$ from our $N_\tau=8$ calculations.

We now are in the position to extract $\mu_B^f$ and $T_f$ from $R_{12}^Q$ and $R_{31}^Q$ which eventually will be measured in the beam energy scan at RHIC \cite{Sahoo, Mitchell}. A large value for $R_{31}^Q$, {\it i.e.} $R_{31}^Q \simeq 2$ would suggest a low freeze-out temperature $T \lesssim 155$ MeV, while a value $R_{31}^Q \simeq 1$ would suggest a large freeze-out temperature, $T \sim 170$ MeV. A value of $R_{31}^Q \simeq 1.5$ would correspond to $T\sim 160$ MeV.
A measurement of $R_{31}^Q$ thus suffices to determine the freeze-out temperature. In the HRG model parametrization of the freeze-out curve \cite{fc} the favorite value for $T^f$ in the beam energy range 200 GeV $\geq \sqrt{s_{AA}} \geq 39$ GeV indeed varies by less than 2 MeV and is about 165 MeV.
At this temperature the values for $R_{31}^Q$ calculated in the HRG model and in QCD differ quite a bit, as is obvious from Fig.~\ref{fig:R31} (right). While $R_{31}^Q \simeq 2$ in the HRG model, one finds $R_{31}^Q \simeq 1.2$ in QCD at $T=165$ MeV. Values close to the HRG value are compatible with QCD calculations only for $T\lesssim 157$ MeV. We thus expect to either find freeze-out temperatures that are about 5\% below HRG model results or values for $R_{31}^Q$ that are significantly smaller than the HRG value. A measurement of this cumulant ratio at RHIC thus will allow to determine $T^f$ and probe the consistency with HRG model predictions.

For any of these temperature values a comparison of an experimental value for $R_{12}^Q$ with Fig.\ref{fig:R31} (middle)  will allow to determine $\mu_B^f$. To be specific, at $T = 160$ MeV we expect to find $\mu_B^f = (20 - 30)$ MeV, if $R_{12}^Q$ lies in the range $0.012 - 0.020$, $\mu_B^f = (50 - 70)$ MeV for $0.032 \leq R_{12}^Q \leq 0.045$ and $\mu_B^f = (80 -  120)$ MeV for $0.05 \leq R_{12}^Q \leq 0.08$. These parameter ranges are expected \cite{fc} to cover the regions relevant for RHIC beam energies $\sqrt{s_{AA}} = 200$ GeV, 62.4 GeV and 39 GeV, respectively. As is evident from Fig.~\ref{fig:R31} (middle) the values for $\mu_B^f$ will shift to smaller (larger) values when $T^f$ turns out to be larger (smaller) than 160 MeV. A more refined analysis of $(T^f , \mu_B^f)$ will become possible, once the ratios $R_{12}^Q$ and $R_{31}^Q$ have been measured experimentally.

\section*{Acknowledgments}
Numerical calculations have been performed on BlueGene computers at the New York Center for Computational Sciences (NY-CCS) at BNL and on clusters of the USQCD collaboration at JLab and FNAL as well as on the GPU-cluster at the University of Bielefeld. We further acknowledge support by contract DE-AC02-98CH10886 with the U.S. Department of Energy, the Bundesministerium f\"ur Bildung und Forschung under grant 06BI9001, the GSI under grant BILAER, the DFG under grant GRK881 and the EU Integrated Infrastructure Initiative Hadron-Physics 3.

\end{document}